\documentclass[twocolumn,showpacs,preprintnumbers,amsmath,amssymb]{revtex4}
\usepackage{graphicx}% Include figure files
\usepackage{dcolumn}% Align table columns on decimal point
\usepackage{bm}% bold math
\usepackage{epsf}
\usepackage{color}

\newcommand{\bcen}{\begin{center}}
\newcommand{\ecen}{\end{center}}
\newcommand{\btab}{\begin{tabular}}
\newcommand{\etab}{\end{tabular}}
\newcommand{\bdes}{\begin{description}}
\newcommand{\edes}{\end{description}}

\newcommand{\beq}{\begin{equation}}
\newcommand{\eeq}{\end{equation}}
\newcommand{\bea}{\begin{eqnarray}}
\newcommand{\eea}{\end{eqnarray}}

\newcommand{\half}{\frac{1}{2}}
\newcommand{\bary}{\begin{array}}
\newcommand{\eary}{\end{array}}
\newcommand{\benum}{\begin{enumerate}}
\newcommand{\eenum}{\end{enumerate}}
\newcommand{\bitem}{\begin{itemize}}
\newcommand{\eitem}{\end{itemize}}
\newcommand{\be} { \mbox{\boldmath $e$}}

\newcommand{\bk} { \mbox{\boldmath $k$}}

\newcommand{\bn} { \mbox{\boldmath $n$}}

\newcommand{\bq} { \mbox{\boldmath $q$}}
\newcommand{\br} { \mbox{\boldmath $r$}}
\newcommand{\bs} { \mbox{\boldmath $s$}}

\newcommand{\bS} { \mbox{\boldmath $S$}}

\newcommand{\mean}[1]{\langle #1 \rangle}

\newcommand{\prn}[1] {(\ref{#1})}

\newcommand{\fig}[1]{fig.~\ref{#1}}
\newcommand{\Fig}[1]{Fig.~\ref{#1}}

\newcommand{\citebyname}[1]{\citeauthor{#1}\cite{#1}}

\newcommand{\figwidth}{9.0truecm}
\newcommand{\nemQ}{{\cal Q}}

\begin{document}

\preprint{}

\title{Possible ferro-spin nematic order in NiGa$_2$S$_4$}
\author{Subhro Bhattacharjee$^{1}$}\email{subhro@physics.iisc.ernet.in}
\author{ Vijay B.~Shenoy$^{2,1}$}\email{shenoy@mrc.iisc.ernet.in} 
\author{T.~Senthil$^{1,3}$}\email{senthil@physics.iisc.ernet.in}
\affiliation{$^{1}$Center for Condensed Matter Theory,   Indian Institute of Science, Bangalore 560 012, India\\
$^{2}$Material Research Center, Indian Institute of Science, Bangalore 560 012, India \\
$^{3}$Massachusetts Institute of Technology, Cambridge, MA 02139, USA. }

\date{\today}

\begin{abstract}
We explore the possibility that the spin-$1$ triangular lattice magnet
NiGa$_2$S$_4$ may have a ferro-nematic ground state with no frozen
magnetic moment but a uniform quadrupole moment. Such a state may be
stabilized by biquadratic spin interactions. We describe the physical
properties of this state and suggest experiments to help verify this
proposal. We also contrast this state with a `non-collinear' nematic
state proposed earlier by \citeauthor{Tsunetsugu2005} for
NiGa$_2$S$_4$.
\end{abstract}

\pacs{75.10.Jm, 75.90.+w}

\maketitle

\section{Introduction}
\label{intro} Spins on a lattice often develop some periodic order
at low temperatures. It has been appreciated for some time that
such ordering may be killed even at zero temperature by quantum
fluctuations. This is particularly true of spin systems in low
dimension and/or the presence of geometric frustration of the
magnetic interactions.  Indeed one of the earliest suggestion of a
``resonating valance bond'' spin liquid state was for the nearest
neighbor antiferromagnetic Heisenberg on the triangular
lattice.\cite{Anderson1973} Although, the ground state of this
system is now believed to be the ``120$^\circ$ ordered'' state
(see \citebyname{Capriotti1999} and references therein), the
experimental search for low spin quantum magnets on a frustrated
perfect triangular lattice has continued.

Recently, \citebyname{Nakatsuji2005} reported an experimental
realization of an insulating spin-$1$ quantum magnet on a perfect
triangular lattice in the chalcogenide NiGa$_2$S$_4$. This is a
layered material with alternating Ni-S and Ga-S planes. Each Ni$^{2+}$
ion is in a (distorted) octahedral environment of six S$^{2-}$ ions,
and the Ni layer has Ni$^{2+}$ ions arranged in a perfect triangular
lattice. The system is an insulator; the only important electronic
degree of freedom being the spin on the Ni$^{2+}$ ion which has an
electronic configuration of $t^6_{2g} e^2_g$ (due to the crystal field
splitting brought about by the octahedral environment of the S$^{2-}$
ions), and due to the Hund coupling in the $e_g$ sector, has
$S=1$. The compound shows unusual magnetic properties. The high
temperature susceptibility is Curie-Weiss like (with a Wiess
temperature of $-$80 K), consistent with an $S=1$ moment on the
Ni$^{2+}$ ions. However at around the Weiss temperature, the
susceptibility begins to drop smoothly, and reaches a finite value at
$T = 0$ indicating the absence of a spin gap. The magnetic specific
heat shows two humps, one at around 10 Kelvin, and another broad hump
at about 80 Kelvin (the Weiss temperature). However there is no sign
of any singular or discontinuous behavior suggesting a phase
transition. The magnetic entropy per spin shows a plateau between 15
to 50 Kelvin at about a third of the high temperature entropy
suggesting large degeneracy of low lying excitations above the ground
state. The specific heat, up to 10 K, shows a $T^2$ power law
suggesting the presence of linear dispersing gapless modes in the Ni
planes.  Moreover, this behavior of the specific heat was found to be
robust to non-magnetic Zn substation in the
compound\cite{Nambu2005}. Intriguingly despite this no periodic spin
order is detected in powder neutron scattering, and a broad (possibly
incommensurate) peak is seen with a wavelength that is roughly twice
that of the well known 120$^\circ$ spin order on the antiferromagnetic
triangular lattice. Furthermore there is no indication of any lattice
distortion associated with possible development of spin-Peierls order
on cooling to low temperature.

Motivated by these experiments, \citebyname{Tsunetsugu2005} have
recently proposed a spin nematic ground state for NiGa$_2$S$_4$.  In
such a state there is no average magnetic moment ($\mean{\bS} = {\bf
0}$) but there is a non-zero quadrupole moment characterized by a
non-zero average of
\bea 
\nemQ^{\alpha \beta}_i
= \half \left( S^\alpha_i S^\beta_i + S^\beta_i S^\alpha_i \right)
- \frac{2}{3} \delta^{\alpha \beta} \label{Nem} 
\eea 
where $S^\alpha_i$ is the $\alpha$-spin component operator at site
$i$, $\delta^{\alpha \beta}$ is the Kroneker delta symbol. In the
state proposed in \cite{Tsunetsugu2005}, the quadrupole order
parameter has a three sublattice structure. Precisely if one writes
\bea
\mean{\nemQ_{\alpha\beta}} = q \left(n^\alpha n^\beta - \frac{1}{3}\delta^{\alpha\beta}\right)
\eea
then the ``director" $\bn$ points along three orthogonal directions in the three sublattices of the triangular lattice. 

In this paper we propose an alternate spin nematic state as a possible
ground state of NiGa$_2$S$_4$.  The particular state we propose may be
dubbed a `ferro-nematic' in the sense that the average quadrupole
moment is spatially uniform in the ground state. The director vector
$\bn$ is independent of site $i$. This is thus distinct from the
Tsunetsugu-Arikawa (TA) state\cite{Tsunetsugu2005} which may be dubbed
a non-collinear nematic.  We discuss a number of different properties
of the ferro-nematic state that makes it attractive as an explanation
of the properties of NiGa$_2$S$_4$. We also point out differences with
the TA state that may be used to distinguish them in experiments.
 
Microscopically the simplest nearest neighbor antiferromagnetic
Hamiltonian for a spin-$1$ triangular magnet is expected to have a
120$^\circ$ spin order which is not seen in NiGa$_2$S$_4$. Thus the
Hamiltonian must apparently involve other terms that destabilize
the magnetic order. This is further supported by the observation
that {\em bulk} NiS$_2$ is close to a Mott transition and hence
may have significant charge fluctuations even when insulating. In
NiGa$_2$S$_4$ the magnetic layers consist of NiS$_2$ sheets - an
effective spin-only description of these may then plausibly have
sizable interactions beyond the simplest near neighbor
antiferromagnetic exchange.

Motivated by these considerations we will discuss the possibility
of spin nematic order in the framework of the model Hamiltonian
\bea 
H & = & H_0 + H_a \\
H_0 & = &  J \sum_{\mean{i,j}} \bS_i \cdot \bS_j - K
\sum_{\mean{i,j}} (\bS_i \cdot \bS_j)^2 
\label{BQHam} 
\eea 
Here $H_0$ is the part of the Hamiltonian that is isotropic in spin
space. $H_a$ refers to small anisotropy terms which may be important
in pinning any possible long range order.  $\bS_i$ is the spin-$1$
operator at the site $i$ of a two dimensional triangular lattice, and
$J, K \ge 0$. Biquadratic interactions of some strength $K$ will in
general be present in spin-$1$ magnets and are known to favor nematic
ordering.\cite{Chandra1991,Podolsky2005} In the present case where the
bilinear exchange $J$ is frustrated the effects of the biquadratic
exchange are somewhat enhanced.  We study the ground state of such a
Hamiltonian in the mean field approximation and construct a zero
temperature phase diagram. We find that at values larger that a
critical value of the bi-quadratic exchange $K/J \sim 1.15$, there is
a first order transition from a magnetic(120$^\circ$ order) state to a
ferro-nematic ordered state. It is the latter that we propose to be
ground state of the NiGa$_2$S$_4$. We then study the expected
properties of the ferro-nematic by including fluctuations beyond the
mean field.

TA also invoked biquadratic interactions to stabilize their
non-collinear nematic state.  However, they chose $K < 0$. Microscopically
a biquadratic term may arise from a few different sources. First in an
underlying two-band Hubbard type description of the $e_g$ orbitals,
the spin Hamiltonian arises describes the gain of kinetic energy due
to virtual fluctuations in a Mott insulating state.  In this large-$U$
perturbation theory the leading term is just the usual
antiferromagnetic exchange. At higher orders a variety of terms will
be induced which include the biquadratic exchange. The sign is readily
guessed by just considering two sites. In a total spin-$2$
configuration of the two sites, there can be no net gain of kinetic
energy -- so $K$ must be such as to make total spin-$2$ a high energy
state. This requires positive $K$. In some systems biquadratic
interactions may also be induced due to spin-phonon coupling. Indeed
if an optical phonon couples to the spin bilinear in a bond, its
effects can be approximately accounted for by integrating it out in
favor of a biquadratic term. Again this leads to a positive $K$. We
note that positive $K$ tends to favor the ferro-nematic state.

\section{Mean field and fluctuations}
\label{MFT}

We now proceed to obtain an approximate ground state of the isotropic
part of the Hamiltonian $H_0$ adopting a mean field approach.  Using
the definition \prn{Nem}, the Hamiltonian $H_0$ can be rewritten as
\bea
 H_0 = (J + \frac{K}{2}) \sum_{\mean{ij}} \bS_i \cdot \bS_j   - K \sum_{\mean{ij}} {\cal Q}^{\alpha \beta}_i {\cal Q}^{\alpha \beta}_j, \label{NemHam}
\eea
ignoring constant terms. The $K$ term is thereby seen to favor
ferro-nematic order.  For small $K$ we expect that the stable ground
state will have spiral magnetic order which may give way to the
ferro-nematic state as $K$ is increased. We therefore calculate the
mean field energies of spiral magnetic and ferro-nematic states to
obtain the mean field phase diagram.

We introduce the expectation values of the spin and nematic operators
respectively as $\mean{\bS_i}$ and $\mean{\nemQ^{\alpha \beta}_i}$ at
each site and obtain the local Hamiltonian at site $i$ as
\begin{widetext}
\begin{equation}
H_{MF}^i=(J+K/2)\sum_\delta ({\bS_i \cdot \mean{\bS_{i+\delta}}}- \half {\mean{\bS_i}} \cdot \mean{ \bS_{i+\delta} })
-K \sum_\delta(\nemQ_i^{\alpha \beta} \mean{\nemQ_{i+\delta}^{\alpha \beta}}
- \half \mean{ \nemQ_{i}^{\alpha \beta}}
\mean{\nemQ_{i+\delta}^{\alpha \beta}})
\label{NemMF}
\end{equation}
\end{widetext}
where $\delta$ runs over all the neighbors of the site $i$.
 
For the ferro-nematic state the mean spin at any site is zero, and the
$\mean{\nemQ^{\alpha \beta}}_i$ has the form
\bea
\mean{\nemQ^{\alpha \beta}}_i = q\left(n^\alpha n^\beta -\frac{1}{3}\delta^{\alpha \beta} \right)
\eea
with $\bn$ independent of the site index $i$. The mean field value of
$q$ is readily calculated to be $-1$ corresponding to an energy per
spin $-2K$.  This energy must be compared with that of the spiral
state.
 
For the spiral state the mean spin at the site $i$ is taken to be of the form
\bea
\mean{\bS_i} = m \bs_i
\eea
with $\bs_i$ a unit vector defined at the site $i$ with position vector $\br_i$  as
\bea
\bs_i = \cos{(\bq\cdot\br_i)} \, \be_x + \sin{(\bq \cdot \br_i)} \, \be_y \label{seq}
\eea
where $m$ and $\bq$ are, as yet undetermined, magnitude and wave
vector respectively, and $\be$s are two orthogonal basis vectors in
the plane of spiral ordering.The spin ordering will induce a non-zero
$\mean{\nemQ^{\alpha \beta}_i}$ of the form
\bea
\mean{\nemQ^{\alpha \beta}_i} = Q \left( s^\alpha_i s^\beta_i - \frac{1}{3} \delta^{\alpha \beta} \right)
\eea
with $Q$ satisfying $\frac{3}{2} m - 1 \le Q \le 1$. 
The four variational parameters $m, q_x, q_y, Q$ are determined by
minimizing the ground state energy of the mean field Hamiltonian
\prn{NemMF}.

\begin{figure}
\centerline{\epsfxsize=8.5truecm \epsfbox{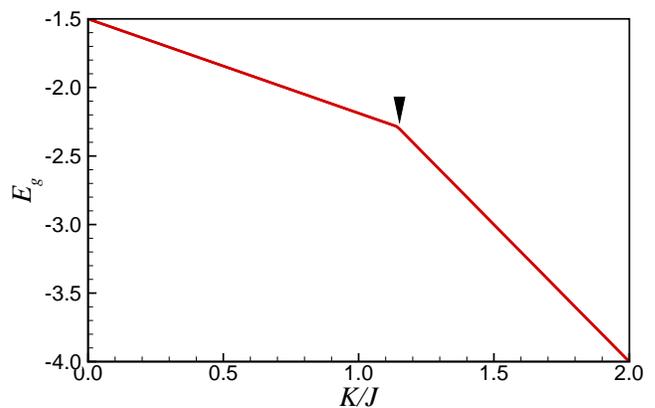}}
\centerline{(a)}
\centerline{\epsfxsize=\figwidth \epsfbox{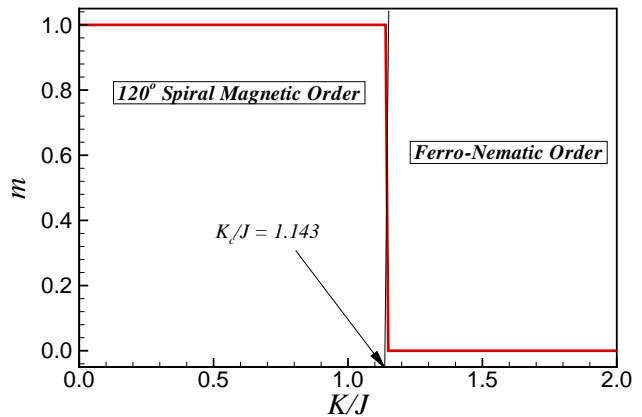}}
\centerline{(b)}
\centerline{\epsfxsize=\figwidth \epsfbox{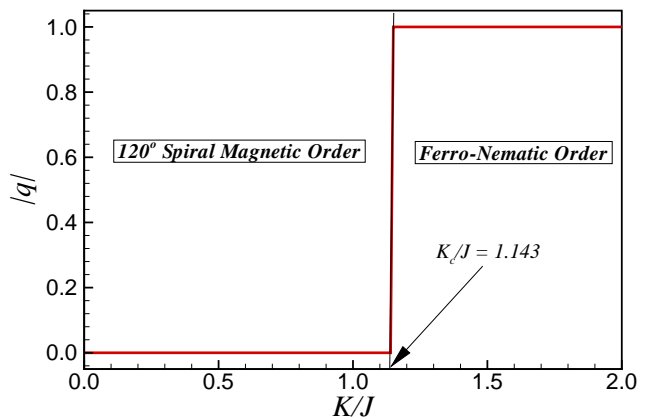}}
\centerline{(c)}
\caption{ (color online) (a) Variation of the mean field ground state energy $E_g$ per site with $K/J$. Note the ``level crossing'' at $(K/J)_c = 1.143$. (b) The variation of $m$ with $K/J$. A first order phase transition occurs at $(K/J)_c = 1.143$ at which $m$ vanishes. (c) The variation of $q$ with $K/J$. There is a jump in $q$ at $(K/J) = 0^+$, and another at the critical value of $(K/J)_c = 1.143$.}
\label{Egmq}
\end{figure}

\Fig{Egmq}(a) shows the variation of the ground state energy as a
function of $K/J$. The standard spiral state with 120$^\circ$ three
sublattice order is obtained as the ground state for $ (K/J)_c \leq
\frac{24}{21} \sim 1.143$. Beyond this value the ferro-nematic state
is found to have lower energy. A study of the spin order parameter $m$
(\fig{Egmq}(b)) shows that it jumps at the transition point indicating
a first order transition.

The structure of the fluctuations beyond the mean field in the spiral
state are well-known and consist of three gapless spin wave modes.
The ferro-nematic state also breaks global spin $SU(2)$ symmetry
despite the absence of a frozen magnetic moment. Indeed the director
$\bn$ picks out a single direction in spin space. Thus this state will
have gapless Goldstone modes corresponding to slow transverse
fluctuations of the director.  A straighforward calculation shows the
existence of two such gapless modes (coresponding to the two
independent transverse directions in which the director can tilt)
centered at wavevector $\bk = {\bf 0}$.  At low energies the
dispersion is linear $\omega = c|\bk|$ with velocity
\bea
c = \frac{1}{\hbar} \sqrt{12 K(J+ \frac{K}{2} )}
\eea

\section{Physical properties of the ferro-nematic state}

We now consider various physical properties of the ferro-nematic state
with an eye toward interpreting experiments in NiGa$_2$S$_4$. In
thinking about experiments it is important to allow for the
possibility of small spin anisotropies that pin the nematic order
parameter. The most important of these is single ion anisotropy that
selects out a particular direction in spin space. Specifically
consider
\bea
H_a = D \sum_i \left(S^z_i\right)^2
\eea
with $D > 0$. The natural choice is to have the hard axis point
perpendicular to the Ni-S layers. Note that this anisotropy couples
directly to the ferro-nematic order parameter. Thus a non-zero $D$
even if small will generally pin the director to point perpendicular
to the layers.  Then in the ferro-nematic state even though there is
no magnetic order the spins will predominantly fluctuate parallel to
the layers at low temperature.

In the presence of a non-zero $D$ the linear dispersion of the small
$|\bk|$ director fluctuation modes will be cut-off - instead a gap of
order $D$ will develop as $\bk \rightarrow 0$. For the specific heat
this implies that
\begin{equation}
C \sim T^2
\end{equation}
so long as $T > T_D \sim D$. In NiGa$_2$S$_4$, precisely this behavior
is seen in the range $0.35$ K $< T <$ 10 K. Thus the ferro-nematic
state can explain the low temperature specific heat provided $D$ is
smaller than about $0.35$ K.

Let us now consider the magnetic susceptibility.  In a single domain
sample assuming that the director is fixed the pure ferro-nematic
state has strongly anisotropic magnetic susceptibility.  The system is
spin-gapped for magnetic fields aligned precisely along the nematic
director, but has finite susceptibility for fields applied
perpendicular to the director.  A straightforward calculation shows
that the susceptibility tensor (per site) is
\bea
\chi^{\alpha \beta} = \frac{g^2 \mu_B^2}{3 K} (\delta^{\alpha \beta} - n^{\alpha}n^{ \beta})
\eea
where $n^{\alpha}$ is the direction of the nematic director (now
ferro-aligned at all sites), $g$ is the gyromagnetic ratio, and
$\mu_B$ is the Bohr magneton. If the director is not fixed however and
is free to rotate it would prefer to orient itself in a direction
perpendicular to the external field, and the susceptibility would be a
constant. In the real system the detailed behavior therefore depends
sensitively on the competition between the pinning due to the spin
anisotropy $D$ and the reorientation energy in a field.  At very low
fields the pinning will win and the susceptibility will be
anisotropic. At higher fields the pinning will be overcome and the
susceptibility will be a constant independent of the field
orientation. As the pinning energy per site $\sim D$ while the
reorientation energy $\sim B^2$ the crossover occurs at a field
strength $\displaystyle{B_D \sim \sqrt{\frac{D}{\chi}} } \sim \sqrt{K D}$. 
Therefore the energy scale associated with the depinning field $B_D$  is much larger than $D$ since $K \gg D$ ($K$ is expected to be of the order of 100 K and $D < 0.35$ K), and thus $B_D$ expected to be a measurably large field. 

In a single crystal sample, the susceptibility at a measuring field $B
\ll B_D$ will therefore be strongly anisotropic while at higher fields
a constant susceptibility independent of field orientation will be
obtained. In polycrystalline samples (where the existing experiments
have been done), due to the existence of domains with differing
orientations of the pining axis, the susceptibility will be
independent of field orientation even at low fields ($< B_D$). However
as the field increases beyond $B_D$ domains with pinning direction
parallel to the field will reorient their directors thereby increasing
the net magnetization. We therefore expect significant increase in the
magnetization on field strengths on the scale of $B_D$.

The constant low-$T$ susceptibility and the $T^2$ specific heat of the
ferro-nematic are consistent with the experiments. However these
properties are also shared by the TA state. What experiments may
distinguish the two states? A useful and direct signature is the
polarization of the spin fluctuation spectrum. In the ferro-nematic
state (with director perpendicular to the Ni-S planes) the spins
should primarily fluctuate in the directions along the planes at low
temperature.  In the TA state on the other hand there should be no
such strong preference for the spins to lie along the planes. Thus
spin polarized neutron scattering experiments on single crystals may
be able to determine which (if either) of these two spin nematic
states is realized in NiGa$_2$S$_4$.  Another useful experiment that
could distinguish the ferro-nematic state from the TA state is the
direction dependent susceptibility measurement in single crystals. As
noted above a vanishing susceptibility is expected for a field smaller
than $B_D$ along the pinning direction in the ferro-nematic state,
with a significant increase in the susceptibility when the field
exceeds $B_D$. Field in  a direction perpendicular to the pinning
direction would obtain a constant susceptibility independent of the
field strength over field scales of order $B_D$ and above.  On the
other hand, in the TA state the anisotropy in the susceptibility is
expected to be much smaller for small fields, and the susceptibility
is expected to remain constant with increasing field.

\section{Conclusion}
\label{conclusion}

In this paper we have proposed that NiGa$_2$S$_4$ may have a
ferro-nematic ground state characterized by a uniform quadrupole
moment. We discussed some of the experimental properties of this state
and suggested experiments to distinguish it from the alternate proposal
of \citebyname{Tsunetsugu2005}. In a simple model for a spin-$1$
triangular magnet with bilinear and biquadratic terms, the latter
promotes nematic order. We suggest that the naturally expected sign
for the biquadratic term prefers ferro-nematic order and not the
non-collinear nematic.  We note that while many properties of either
spin nematic state (specific heat, susceptibility) resemble that seen
in experiments neither proposal directly addresses the short ranged
incommensurate spin fluctuations inferred from powder neutron
data. This feature of the spin fluctuation spectrum presumably depends
on the details of the microscopic spin Hamiltonian which are not known
at present.

Just prior to submission of this paper cond-mat/0605234
\cite{Lauchli2006} appeared which also studies the ferro-nematic
state in the J-K spin-$1$ triangular magnet, and suggests this as a
possible ground state in NiGa$_2$S$_4$.

VBS acknowledges support by Indian National Science Academy through
the Young Scientist Programme.  TS was generously supported by a
DAE-SRC Outstanding Investigator Award in India. The authors thank
Leon Balents, Kedar Damle, Matthew Fisher, T.V. Ramakrishnan, and
Diptiman Sen for discussions.

\bibliography{nem}

\end{document}